\documentclass{article}

\newtheorem{Theorem}{Theorem}[section]
\newtheorem{Lemma}{Lemma}[section]
\newtheorem{Proof}{Proof}[section]

\usepackage{amsmath}
\usepackage{graphicx}
\usepackage{enumerate}
\usepackage{natbib}
\usepackage{url} 
\usepackage{amssymb}
\usepackage{graphicx}
\usepackage{multirow}
\usepackage{dsfont}
\usepackage{booktabs}
\newbox\keywdbox
\def\keywords{\global\setbox\keywdbox\vbox\bgroup\hsize\textwidth\small\leftskip0pc\rightskip\leftskip\noindent{\sc
Key words:\hskip1em}\ignorespaces}
\def\endkeywords{\egroup}

\title{A Novel Statistical Test for Treatment Differences in Clinical Trials using a Response Adaptive Forward Looking Gittins Index Rule}

\author{Helen Yvette Barnett$^{1,*}$ \textit{h.barnett@lancaster.ac.uk}, \\
Sof{\'i}a S Villar$^{2}$, Helena Geys$^{3}$ and Thomas Jaki$^{1,2}$ \\
$^{1}$Medical and Pharmaceutical Statistics Research Unit,\\ Lancaster University, United Kingdom \\
$^{2}$MRC Biostatistics Unit, University of Cambridge, United Kingdom \\
$^{3}$Janssen Pharmaceutica, Beerse, Belgium. }
\date{}

\begin{document}

\label{firstpage}

\begin{abstract}
The most common objective for response adaptive clinical trials is to seek to ensure that patients within a trial have a high chance of receiving the best treatment available by altering the chance of allocation on the basis of accumulating data. Approaches which yield good patient benefit properties suffer from low power from a frequentist perspective when testing for a treatment difference at the end of the study due to the high imbalance in treatment allocations. In this work we develop an alternative pairwise test for treatment difference on the basis of allocation probabilities of the covariate-adjusted response-adaptive randomization with forward looking Gittins index rule (CARA-FLGI). The performance of the novel test is evaluated in simulations for two-armed studies and then its applications to multi-armed studies is illustrated. The proposed test has markedly improved power over the traditional Fisher exact test when this class of non-myopic response adaptation is used. We also find that the test's power is close to the power of a Fisher exact test under equal randomization.
\end{abstract}


\maketitle

\section{Introduction}
Equal randomization between treatment arms is the gold standard for any clinical trial \citep[e.g.][]{Shulz1996}, as such a randomization scheme will give the trial the highest power to detect a treatment difference \citep{Pocock1979} under certain assumptions. Whilst the purpose of any trial is to gain information about an experimental treatment, there is also the ethical consideration of the patients in the trial, and these two goals often conflict with one another. This has triggered the development of sequential trial designs, where the probability of a patient receiving a particular treatment is altered sequentially throughout the trial based on previous patients' responses in order to treat subsequent patients on treatments that are believed to be superior. The use of such adaptive randomization has long been suggested for implementation in clinical trials for the advantages in patient benefit it offers \citep[e.g.][]{Rosenberger1993, Bai2002,Hu2006} and a vast methodological literature has been developed on the subject of how to update this patient allocation rule \citep[e.g.][]{Rosenberger2016, Williamson2017}.

Multi-armed bandit models are the optimal idealised solution (in terms of patient benefit) to response adaptive allocation \citep{Gittins1979}. While their original motivation was within trials, they have found wide application outside of trials \citep[e.g.][]{Gittins2011,Vermorel2005} but have, to our knowledge, not actually been used in the clinical trials setting. One of the reasons is that both the optimal solution and the computationally efficient approximate solution, the Gittins Index Rule \citep{Gittins1979}, is deterministic. In the clinical trials setting, the deterministic nature of the Gittins Index Rule for patient allocation and the assumption of infinite sample size are problematic due to the inherent risk of bias \citep{Hardwick1991}. Modifications of the Gittins Index Rule have been developed to apply the classic multi-armed bandit framework to deriving nearly-optimal patient allocation procedure in clinical trials using adaptive randomization \citep{Villar2015, Williamson2020}. These modifications allow for randomization, consider finite-sized trials and cater for patient accrual in blocks rather than individually.

One particular modification of note is known as the modified Forward Looking Gittins Index Rule \citep{Villar2015b} which offers the advantage that patient allocation is no longer deterministic. Instead, the patient allocation is random according to an allocation probability. This probability can be calculated exactly or using Monte-Carlo simulations that themselves use the deterministic Gittins Index Rule.

Extending to adaptive randomization to adjust allocation probabilities according to covariates of patients is an important step in the area of personalised medicine. Adjusting for covariates or biomarkers not only allows for higher levels of patient benefit within the trial, but also for the targeting of experimental treatments to patient groups that will see the most rewards when the treatment is marketed. \cite{Villar2018} introduce a method that allows covariate adjustment, using a modified forward looking Gittins Index rule, henceforth referred to as the CARA FLGI procedure.

Non-myopic bandit-based procedures offer much in the way of patient benefit. The process of calculation of the patient allocation probabilities allows for looking forward a considerable way into future patients' allocations. Hence they do not have the shortcomings associated with myopic procedures that only take into account past information in the allocation of the present patient, namely that in the exploring vs exploiting theory of allocation rules \citep{Gittins2011}, they do not explore enough. Consequently, such myopic procedures can settle too early on an inferior arm. Like all response adaptive designs, they have the drawback that the resulting designs often lack in power due to substantial imbalance between patient groups. 

The purpose of most response adaptive trials is to identify superior treatments quickly, and in doing so, the resulting patient allocation favours the superior treatment. When directly performing inference on the binary responses of the patients to determine the outcome of the trial, there must be a trade-off between patient benefit and power. For a two-armed trial (under equipoise and equal variance assumptions), the closer allocation is to equal sample sizes in each group, the higher the power, but smaller the patient benefit. Likewise the further from equal the sample sizes between treatments can be, the larger the potential patient benefit but lower the power. Bandit-based designs provide high patient benefit, and therefore suffer from low power, which is a concern for their implementation \citep{Villar2015}.

For bandit-based designs, when the patient allocation favours a treatment, this is an indication of the superiority of that treatment. It is therefore intuitive to use this indication to analyse results from a trial utilising an FLGI procedure. In this paper we discuss how testing based on allocation probabilities from the CARA FLGI procedure can be used as an alternative to testing based on the binary outcome of response to test the null hypothesis of no treatment difference. Alternative approaches to inference that are tailored to the specific response adaptive randomization algorithm, such as the randomization test \citep{Simon2011} have been applied to the FLGI design \citep{Villar2018a}. Such an approach preserves type I error under broad assumptions but it results in substantial power loss when compared to traditional inference in a fixed randomization trial. Motivated by the potential for a higher powered response-adaptive randomized trial, this novel approach to decision making provides an exciting opportunity for alternative methodologies in adaptive randomization. Such an alternative methodology is exactly the solution to one of the key concerns posed in the National Science Foundation 2019 report, \textit{Statistics at a Crossroads: Who is for the Challenge?}, stating ``A fundamental issue is the development of inference methods for post subgroup selection'' \citep{He2019}. 

In the following, Section \ref{sec:prop} provides a review of the CARA FLGI method, in particular how the allocation probabilities are calculated. Section \ref{sec:null_dist} derives the testing procedure for the use of the allocation probabilities in such a trial. Section \ref{sec:results} then illustrates the use of the testing procedure in both a real multi-arm trial scenario and simulations. Alongside this, its comparative properties and advantages over alternative methods are presented before we conclude with a discussion in Section \ref{sec:discussion}.

\section{Properties of Allocation Probabilities} \label{sec:prop}
\subsection{The CARA FLGI Procedure}
Throughout this paper, it is assumed that the CARA FLGI procedure, as described by \cite{Villar2018}, is applied. It is worth noting that although in the following we focus on the covariate-adjusted case, where there are multiple biomarker categories that partition the patient population, the procedure of using allocation probabilities for inference purposes as we describe can indeed be applied to any FLGI allocation such as those presented by \cite{Villar2015b}; this is a strong advantage of the procedure. In fact, since the CARA FLGI procedure reduces to the simpler FLGI allocation when there is only a single biomarker category, we also evaluate this case in the following work. It is also worth emphasizing that what we propose here is a novel testing procedure for class of response adaptive designs, not a novel response adaptive design itself. The testing procedure can be used when interested in comparing an experimental arm (possibly out of many) to a control arm. 

The trial set-up is as follows. Patients are accrued in blocks of pre-specified size $B$, with total trial sample size $N=KB$, where $K$ is the number of blocks. Biomarker categories that partition the patient population are defined, and each patient has an associated biomarker category. At the beginning of each block, an allocation probability is calculated for each biomarker category using the Forward Looking Gittins Index rule. This allocation probability is then used as the probability of assigning a patient within that biomarker category to the experimental treatment. This differs from using the standard Gittins Index rule, as there is still randomness in the patient allocation. The larger the block size, the higher the randomization element, although this comes at the expense of deviating further from the optimality of patient benefit. We consider the one-sided hypothesis to be tested as 

\begin{eqnarray*}
H_0:&&\mbox{ No treatment difference in subgroup $z$ } \\H_A:&&\mbox{ The experimental treatment is superior to the control in subgroup $z$}
\end{eqnarray*}
Note that extensions to two-sided tests are straightforward. Traditionally, frequentist inference is carried out using statistical tests that are based on the observed success/failure outcomes from the trial. We propose however, to use the CARA FLGI allocation probabilities calculated at the beginning of each block to test these hypotheses. In the following, we provide an overview of how these probabilities are calculated using the Forward Looking Gittins Index in order to understand why they can be used for testing the hypotheses.


\subsection{Calculation of CARA FLGI Allocation Probability Distribution} \label{sec:FLGI_exp}

At the beginning of every block in the trial, the CARA FLGI procedure calculates an allocation probability per treatment arm, per biomarker category. This FLGI procedure can use allocation probabilities calculated exactly (\citep{Villar2015b}), however the theoretical calculation is extremely intensive and therefore Monte Carlo simulations are often used to calculate these probabilities. In the following, we assume the use of Monte Carlo simulations.  For simplicity, we here assume that there are two treatment arms, labelled 0 for control and 1 for experimental, although the calculations are identical if multiple treatment arms are used. Consider biomarker categories labelled  $z=1,\ldots n_{z}<\infty$ and let us recap how the procedure calculates the allocation probability for the experimental treatment for biomarker category $z$,  $p_{\mbox{{\small alpro, z}}}$, using Monte Carlo simulations. It is worth noting that the following calculations are neither under the null or alternative hypotheses as there is no assumption on treatment difference when calculating the probabilities themselves. 

First, consider the current states of all biomarker categories at the beginning of the block, which are defined by the number of successes on the standard treatment ($s_{0,z}$); failures on the standard treatment ($f_{0,z}$); successes on the experimental treatment ($s_{1,z}$) and failures on the experimental treatment ($f_{1,z}$). We denote the current state $i$ in category $z$ by $S^i_{z}(s^i_{0,z},f^i_{0,z},s^i_{1,z},f^i_{1,z})$ and the starting state for the block by  $S^1_{z}(s^1_{0,z}, f^1_{0,z}, s^1_{1,z}, f^1_{1,z})$.  For the first block in the trial, this state is specified via an uninformative prior of $S_z^1(1,1,1,1)$. From each of these category states, the procedure takes $n$ Monte Carlo runs labelled $j=1, \ldots , n$; each run is an independent block of the pre-specified block size $B$. 

Within each Monte Carlo run, $j$, the first patient is allocated to one of the treatment arms and success/failure is observed. The state for that biomarker category is updated, and the next patient is allocated to a treatment arm based on their (possibly updated) biomarker category state. This continues until the block is full, noting that patients in the same block may have different biomarker categories. This is repeated for each run, starting at the same initial states for each category.

The allocation of patients in the FLGI allocation procedure (and therefore in the Monte Carlo simulations) depends on the Gittins Index ($GI$) Rule \citep{Gittins1979}. For a given patient, this rule takes the two available treatment arms (standard and experimental for the patient's biomarker category) and calculates the $GI$ for each arm. The patient is allocated to the treatment arm with the highest $GI$, breaking ties at random. At any given point, and consistent with the multi-armed bandit framework, we assume that patient success for a given treatment arm occurs according to the posterior success probability so far on that treatment arm.

When FLGI probabilities are estimated through a Monte Carlo procedure, the allocation probability for each category is the proportion of patients in each category allocated to the experimental treatment over the total number of runs.

This allocation probability is calculated (or approximated for large blocks) as:
\begin{equation}
p_{\mbox{{\small alpro, z}}}=\frac{\sum_{j=1}^{n} Y_{j,z} }{\sum_{j=1}^{n} X_{j,z}}. \label{eq:allo_def}
\end{equation}

Assuming that every biomarker category is equally likely to be observed in each block, $X_{j,z} \sim \mbox{Bin} (B, \frac{1}{n_{z}})$ is the number of patients belonging to biomarker category $z$ (regardless of treatment) in the $j$th Monte Carlo (MC) run and $Y_{j,z}$ is the number of patients in category $z$ allocated to the experimental treatment on the $j$th MC run. The assumption upon the biomarker distribution may be relaxed if required; we include it here for simplicity of further calculations.

In order to calculate the distribution of $p_{\mbox{{\small alpro, z}}}$, we consider the cumulative distribution function $F_{p_{\mbox{{\tiny alpro}}}, z}(c) = \mathds{P}(p_{\mbox{{\small alpro, z}}} \leq c)$, which is equivalent to 

\begin{equation*}
F_{p_{\mbox{{\tiny alpro}}}, z}(c) = \mathds{P} \left( \frac{\sum_{j=1}^{n} Y_{j,z} }{\sum_{j=1}^{n} X_{j,z}} \leq c \right) = \mathds{P} \left( \sum_{j=1}^{n} Y_{j,z}  - c \sum_{j=1}^{n} X_{j,z} \leq 0 \right).
\end{equation*}
 
Note that the above assumes $\sum_{j=1}^{n} X_{j,z}  > 0$. If no patients are in category $z$, ($\sum_{j=1}^{n} X_{j,z}  = 0$), then the allocation probability is taken as 0.

The derivation of the discrete joint distribution of $X_{j,z}$ and $Y_{j,z}$ is provided in Web Appendix B in the online supporting information. For a given $j$, let the expectations be denoted as $\mathds{E}(X_{j,z})=\mu_{x,z}$ and $\mathds{E}(Y_{j,z})=\mu_{y,z}.$ and variances $\mbox{Var}(X_{j,z})=\sigma^2_{x,z}$ and $\mbox{Var}(Y_{j,z})=\sigma^2_{x,z}$. The covariance of $X_{j,z}$ and $Y_{j,z}$ is given as $V_z= \sum_{x_{j,z} , y_{j,z}} P(X_{j,z}=x_{j,z} \hspace{5pt} \& \hspace{5pt} Y_{j,z}=y_{j,z})(x_{j,z} - \mu_{x,z}) (y_{j,z} - \mu_{y,z})$ where the sum is over all possible values of $x_{j,z}$ and $y_{j,z}$. For $j \neq j'$, $\mbox{cov} ( X_{j,z} , Y_{j',z}) = 0$ and therefore $\mbox{cov} (  \sum_{j=1}^{n} Y_{j,z}  , c \sum_{j=1}^{n} X_{j,z})=cnV_z$.

We approximate the following using the Central Limit Theorem:
\[
\sum_{j=1}^{n} Y_{j,z} \sim \mathcal{N} \left(n \mu_{y,z} , n \sigma_{y,z}^2 \right),
\sum_{j=1}^{n} X_{j,z} \sim \mathcal{N} \left(n \mu_{x,z} , n \sigma_{x,z}^2 \right)
\]
and hence obtain the cumulative distribution function
\begin{equation} 
F_{p_{\mbox{{\tiny alpro}}}, z}(c) = \Phi \left( \frac{n^{\frac{1}{2}} (c \mu_{x,z} - \mu_{y,z} )}{\sqrt{ \sigma_{y,z}^2 + c^2  \sigma_{x,z}^2 - 2cV_z}} \right), \label{eq:F_def}
\end{equation}
with density function:
\begin{equation} \label{eq:dist}
f_{p_{\mbox{{\tiny alpro}}}, z}(c) = \frac{n^{\frac{1}{2}}(\mu_{x,z} \sigma_{y,z}^2 + c \mu_{y,z} \sigma_{x,z}^2 - \mu_{y,z} V_z - cV_z\mu_{x,z}) }{(\sigma_{y,z}^2 + c^2 \sigma_{x,z}^2 - 2cV_z)^{\frac{3}{2}}} \times \phi \left( \frac{n^{\frac{1}{2}} ( c \mu_{x,z} - \mu_{y,z} )}{\sqrt{ \sigma_{y,z}^2 + c^2  \sigma_{x,z}^2 - 2cV_z}} \right).
\end{equation}

Thus the distribution has a mode at $\frac{\mu_{y,z}}{\mu_{x,z}}$. For a single block starting at state $S^i_{z}(s^i_{0,z},f^i_{0,z},s^i_{1,z},f^i_{1,z})$, this is the ratio of the expected number of patients allocated to the experimental treatment with biomarker category $z$  compared to the total expected number of patients with biomarker category $z$.

\section{Testing for Superiority with Allocation Probabilities}  \label{sec:null_dist}

We present the following Theorem for testing for superiority using allocation probabilities.

\begin{Theorem} \label{threm}
Denote the true difference in success probability on the experimental treatment and control by $p_1-p_0$. Consistently higher allocation probabilities for the experimental treatment, i.e. more allocation probabilities greater than 0.5 at beginning of blocks within the trial, are observed if and only if $p_1-p_0>0$.
\end{Theorem}

The theorem is a direct result of the following three Lemmas, the proofs of which are given in Web Appendix C in the online supporting information.

\begin{Lemma} \label{mirror}
For any state $S^i_z(s^i_{0,z}, f^i_{0,z}, s^i_{1,z}, f^i_{1,z})$ with $\mathds{P}(p^{t_1}_{\mbox{{\small alpro, z}}}<0.5)= \gamma$, its``mirror'' state $S^i_z(s^i_{1,z}, f^i_{1,z},s^i_{0,z}, f^i_{0,z})$ will give $\mathds{P}(p^{t_1}_{\mbox{{\small alpro, z}}}>0.5)= \gamma$.
\end{Lemma}

\begin{Lemma} \label{b}
Consistently higher allocation probabilities for arm 1 implies $p_1-p_0>0$. 
\end{Lemma}

\begin{Lemma} \label{a1}
$p_1-p_0>0$ implies consistently higher allocation probabilities for arm 1. As the treatment difference gets large enough, the allocation probability for arm 1 will tend to 1.
\end{Lemma}

In order to use the allocation probabilities to test for superiority, we utilize the distribution of the allocation probabilities under the assumption that the treatments are equal (null distribution). The distribution of the allocation probability for a given block described by equations \ref{eq:F_def} and \ref{eq:dist} is calculated solely from the state at which that block starts. The success probabilities used in the CARA FLGI algorithm are the posterior probabilities within the simulation, and not under any assumptions on the treatments themselves. However, the full null distribution of allocation probabilities for any block, $k$, is under the assumption that the treatments are equal in their success probabilities. This distribution is a mixture distribution; a weighted sum of the allocation probability distributions $f_{p_{\mbox{{\tiny alpro}}}, \zeta, z}$ from the potential states $\zeta$, of which there are say $Z$,  at the beginning of block $k$. The weights are the probabilities of being in each of the potential states $\mathds{P}_z(\zeta)$, for a given equal success probability for both treatments.

\begin{equation} \label{eq:mix_dist}
g^{(k)}(c)=\sum_{\zeta=1}^{Z} f_{p_{\mbox{{\tiny alpro}}}, \zeta, z}(c) \mathds{P}_z(\zeta).
\end{equation}

This mixture distribution has point masses at 0 and 1, as for certain extreme states, the probability of allocating a patient to the experimental treatment is either 0 or 1. As this can occur fairly frequently, the resulting distribution can often not be used to formulate a non-randomized level-$\alpha$-test \citep[see Figure 3 in][]{Smith2018}. To overcome this problem we instead consider the number of blocks for which the allocation probability exceeds 0.5. 

To achieve this, the allocation probabilities are dichotomised according to whether they are greater than 0.5. We then denote the binary outcome $\alpha_k$ as 1 if the allocation probability to the experimental arm for block $k$ is greater than 0.5 and 0 otherwise. Our test statistic, $Q=\sum_{k=1}^{K}\alpha_k$, is then the total number of blocks for which the allocation probability to the experimental arm is larger then 0.5. Note that the value of 0.5 is for the two-arm setting. For a trial with multiple arms, this is the reciprocal of the number of arms.

The discrete distribution of $Q$ under the assumption of no treatment difference is given in Web Appendix D in the online supporting information. Using this distribution we can then find the critical value as the smallest value $c_q$ that satisfies $P(Q > c_q | p_0=p_1)<\alpha$ and reject the null hypothesis if $Q > c_q$ as usual.

Note that for larger sample sizes, the calculation of the exact distribution is computationally intensive and it may be more useful in practice to estimate the distribution via Monte Carlo simulation. 

\begin{figure}[ht!]
\begin{center}
\includegraphics[width=1\textwidth]{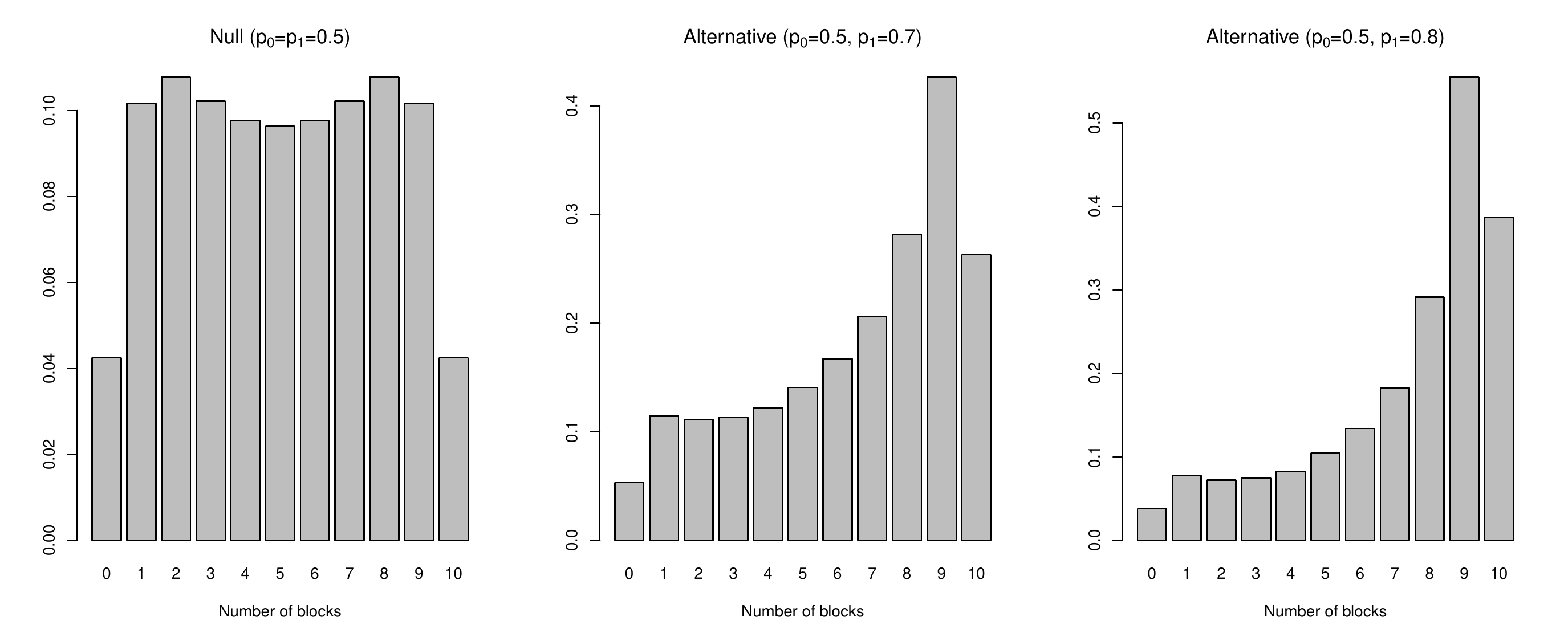}
\end{center}
\caption{An illustration of the null distribution and two alternative distributions of $Q$, the total number of blocks for which the allocation probability to the experimental arm is greater than 0.5, for $K=10$, $B=2$ and $n_z=2$.}
\label{fig:dist20}
\end{figure}

The leftmost plot in Figure \ref{fig:dist20} shows the null distribution of $Q$ for a total sample size of 20, split into 10 blocks of size 2, for $n_{z}=2$. As expected the distribution is symmetric about the midpoint of 5 due to the assumption that treatments have equal success probabilities. In this example distribution, the probability of seeing 10 blocks each with allocation probability to the experimental above 0.5 is 0.043. Hence in order to conclude that there is evidence to suggest the experimental treatment is superior at the 5$\%$ level, we must observe more than 9 allocation probabilities greater than 0.5.

For such a test, the power is not adversely affected by the imbalance in treatment groups. A bigger underlying treatment difference gives more skewed allocation probabilities which means more imbalance between groups (See Figure \ref{fig:dist20}). When performing traditional inference on the outcomes of the trial, for example using a Fisher exact test, the assumptions required for the validity of the test are violated by the heavy dependencies on the outcomes and sampling direction. Therefore the increase in power from a larger treatment difference is lessened by the imbalance in treatment groups. When testing using the allocation probabilities as described above, this is not the case. We have constructed a test that has a structural property of the design embedded into it and therefore it better aligns to the properties of the experiment underlying it.

\section{Application} \label{sec:results}

\subsection{Simulations} \label{sec:simulations}
In order to compare the use of tests based on allocation probabilities versus those based on success rates, we compare and evaluate the performance with simulated datasets. Since we envisage that the main advantage of using allocation probabilities is to increase the power of the test for superiority, we compare the power for treatments with varying success probabilities in simulated two-arm trials that implement the CARA FLGI procedure. 

We compare the results from the proposed procedure to three other methods. The first two use the same adaptive randomization procedure but analyse the results using a) a Fisher exact test of success rates and b) a GLM with $logit$ link function ( $logit(\rho_z)=\beta_0+ \beta_1 T$, where $\rho_z$ is the success rate for biomarker category $z$ and $T$ is an indicator variable taking the value 1 if a patient is assigned to the experimental treatment). randomization test are not compared since they have already been shown to be inferior to the Fisher exact test in terms of power \citep{Villar2018a} in the this class of designs. The third method uses equal randomization between the arms and results are analysed using a Fisher exact test on success rates. These trials will have high power, but much smaller patient benefit than those using adaptive randomization and are used as a benchmark in terms of power. We also report the percentage of patients on the best treatment, and the total number of observed successes. We have intentionally not included comparisons to alternative covariate adjusted response adaptive designs, since our objective is to improve the power of a particular class of designs by using a novel testing procedure, not a novel adaptive design.

Trials of three different sample sizes are considered, $N=40,80,160$. For the CARA FLGI procedure, a block size of $B=2$ is used initially with an extension to $B=4$ and $B=8$ to assess the impact of block size. For the use of allocation probabilities, the first two blocks' allocation probabilities are disregarded as burn-in, as the CARA FLGI procedure must be adequately established before the allocation probabilities are meaningful and can be interpreted. This number is chosen as a balance between the procedure being established (one cannot reasonably expect it to be fully established after two blocks) and using the most information available from the allocation probabilities. For smaller number of blocks this is very important, however the effect is diluted for larger numbers of blocks. In practice this burn-in can be tailored to suit the trial size and block size. It must be noted that the burn-in is for inference only, the sample size includes those patients in the blocks whose allocation probabilities are disregarded as burn-in. For each sample size, the number of biomarker categories considered are $n_{z}=1,2,3,4$.

As it is known that the Fisher exact test can lead to a conservative type I error rate \citep[e.g.][]{Storer1990}, we adjust the rejection criteria in each case to ensure that a 5\% type I error rate is observed in the simulations in order for a fair comparison between methods. For the Fisher exact test and GLM, this is implemented by simply adjusting the critical value for rejection. For the test using allocation probabilities, we use a randomized test. Results that are unadjusted for type I error rate are given in Web Appendix A in the online supporting information, which also show that the proposed procedure controls the type I error rate before adjustment.

\begin{figure}[ht!]
\begin{center}
\includegraphics[width=0.7\textwidth]{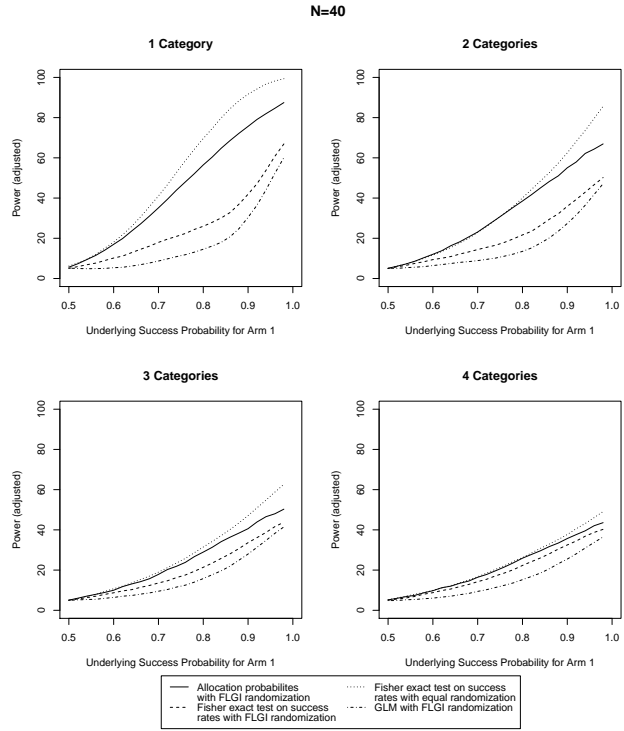}
\end{center}
\caption{Comparison of power for $N=40$ \& $B=2$; rejection criteria adjusted for type I error rate.}
\label{fig:N40a}
\end{figure}


\begin{figure}[ht!]
\begin{center}
\includegraphics[width=0.7\textwidth]{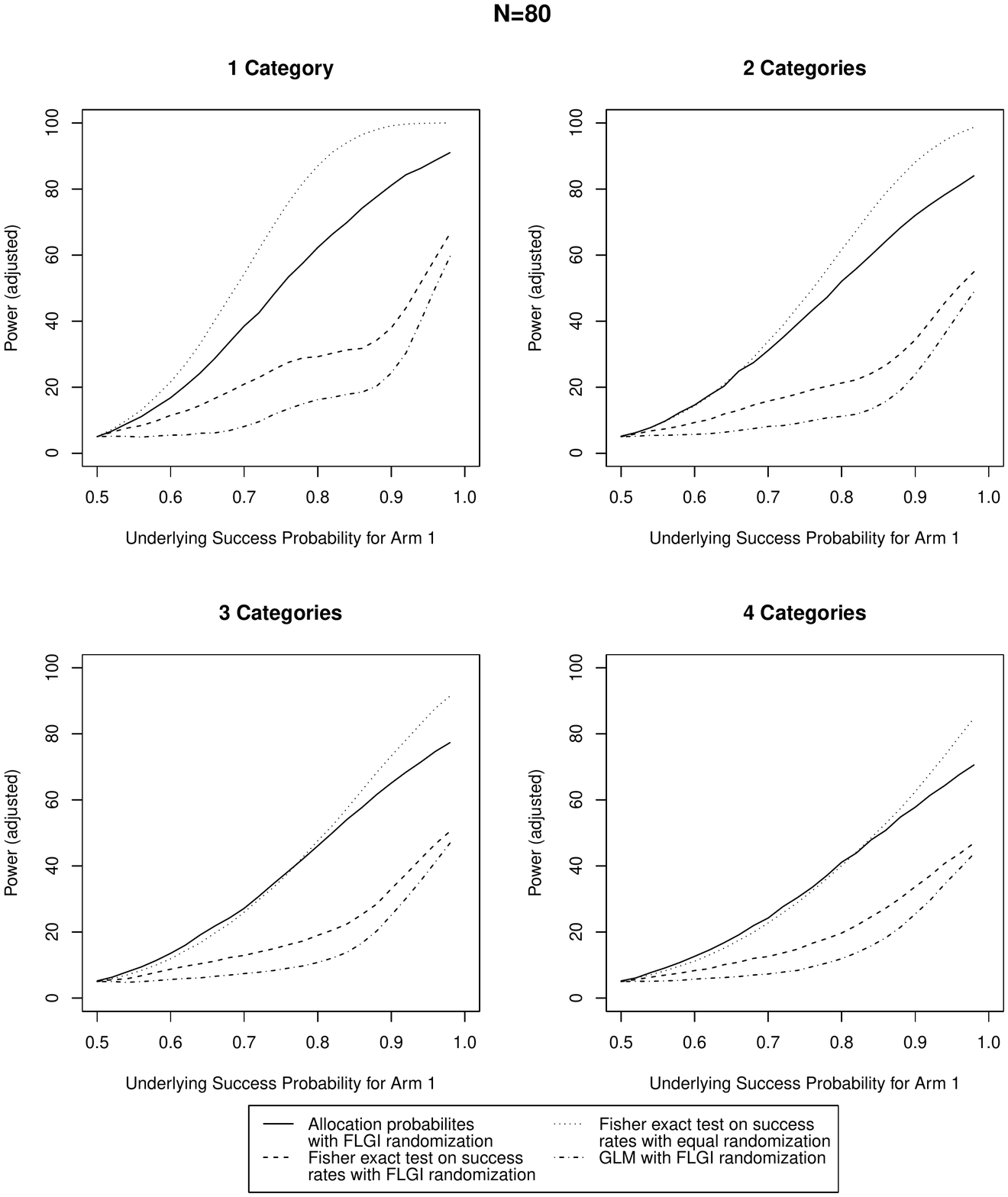}
\end{center}
\caption{Comparison of power for $N=80$ \& $B=2$; rejection criteria adjusted for type I error rate.}
\label{fig:N80a}
\end{figure}


\begin{figure}[ht!]
\begin{center}
\includegraphics[width=0.7\textwidth]{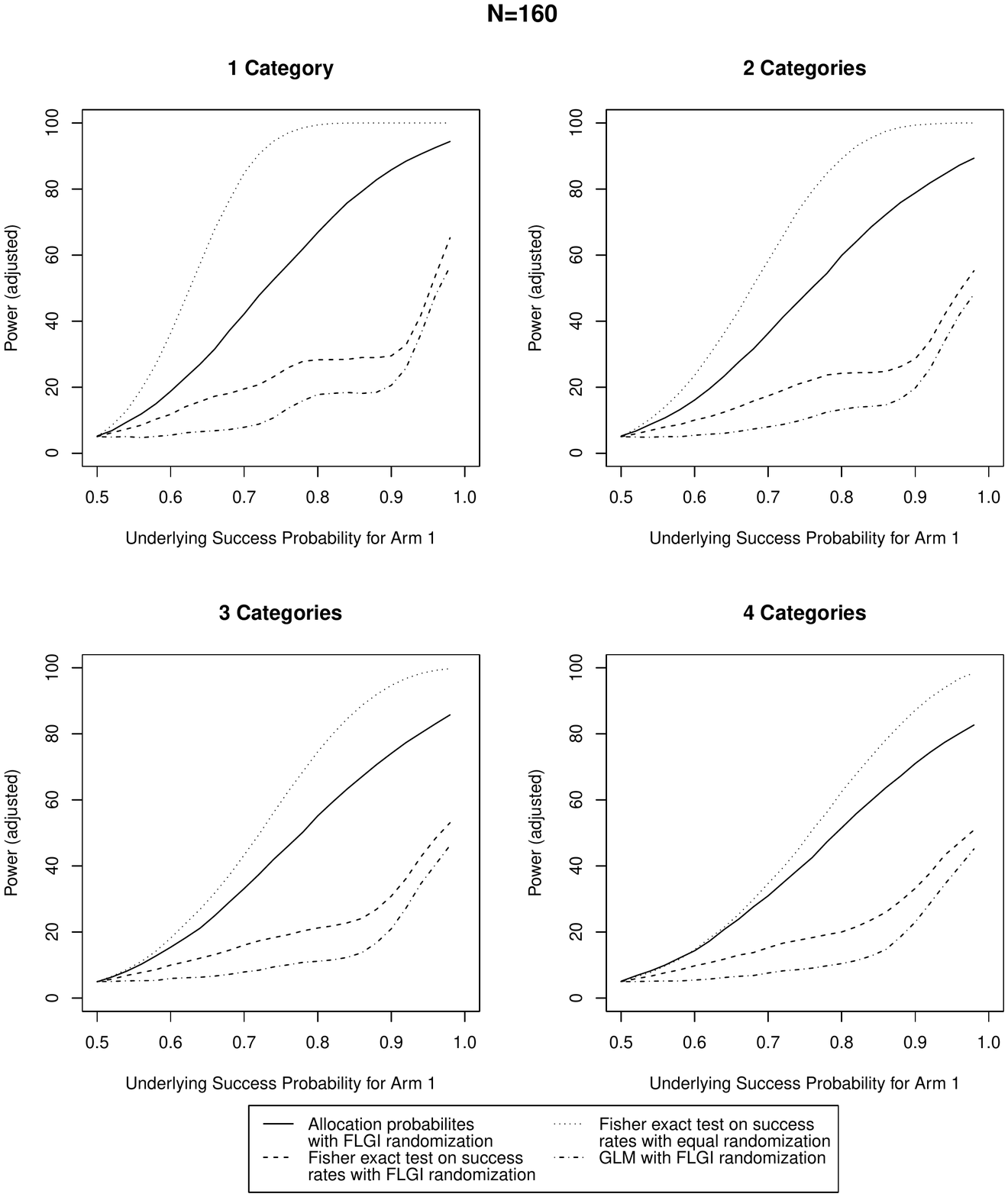}
\end{center}
\caption{Comparison of power for $N=160$ \& $B=2$; rejection criteria adjusted for type I error rate.}
\label{fig:N160a}
\end{figure}


Figures \ref{fig:N40a} - \ref{fig:N160a} compare the power across the 12 scenarios when type I error is adjusted for. In each case, the success probability of the control treatment is set to 0.5 and the success probability of the experimental treatment increases along the $x$-axis. The most notable characteristic of all of these graphs is that the power curve for the procedure using allocation probabilities of the CARA FLGI procedure closely follows the curve for the Fisher exact test using success rates in the equal randomization design. At the same time the power of the test based on allocation probability is markedly higher than the Fisher exact test and the logistic model when the CARA FLGI procedure is used to allocate patients.

The power of the test using allocation probabilities is minimally affected by an increase in the number of categories for the larger sample sizes, whereas the power of the Fisher exact test is adversely affected by an increase in categories. For the scenarios with four biomarker categories, the difference between the Fisher exact test applied to the equal allocation simulations and the use of allocation probabilities applied to the simulations using the CARA FLGI procedure is at most 20\%. Whereas the gain of using allocation probabilities as opposed to the Fisher exact test when the CARA FLGI is used to allocate patients is up to 40\% for the larger sample sizes. The smaller the sample size, the closer the power of the Fisher exact test applied to the equal allocation simulations and the use of allocation probabilities applied to the simulations using the CARA FLGI procedure. Any difference in power is only noticeable in the case of one biomarker category.

When considering the effect of varying block sizes on the power of the procedure, a comparison between the proposed method using allocation probabilities to test for treatment difference and the Fisher exact test on success rates using FLGI randomization is presented in Figure 4 in Web Appendix A in the online supporting information. For larger numbers of categories, the power of the proposed method is well maintained. However, the power is adversely affected for increasing block size when only one category is considered; there is a clear relationship between both the number of categories and block size. Although the power of the Fisher exact test on success rates increases for larger block sizes due to the increased balance between treatment groups (and hence lesser patient benefit), it still does not achieve the power of the proposed procedure with $B=2$.  A larger block size may be advantageous in a trial for practical reasons, but both the largest patient benefit and highest power are achieved using the proposed procedure and a smaller block size. If a larger block size is required, in order to maintain power and patient benefit we recommend $B\leq n_z$. In relation to this, we also recommend a minimum number of blocks of $J \geq 20$.

\begin{table}
\caption{\label{tab:PB20}Percentage of patients on the correct treatment for them, and average total observed successes using CARA FLGI with $B=2$. True treatment difference of 20\%.}
\begin{tabular}{@{}lllllllll@{}}
\toprule
               & \multicolumn{4}{c}{\textbf{Patients on Correct Treatment}}                                    & \multicolumn{4}{c}{\textbf{Average Total Successes}}                                          \\ \midrule
               & \textbf{$n_z$=1} & \textbf{$n_z$=2} & \textbf{$n_z$=3} & \textbf{$n_z$=4} & \textbf{$n_z$=1} & \textbf{$n_z$=2} & \textbf{$n_z$=3} & \textbf{$n_z$=4} \\
\textbf{N=40}  & 77\%                  & 71\%                  & 67\%                  & 65\%                  & 26                    & 26                    & 25                    & 25                    \\
\textbf{N=80}  & 85\%                  & 78\%                  & 74\%                  & 72\%                  & 53                    & 52                    & 52                    & 51                    \\
\textbf{N=160} & 90\%                  & 84\%                  & 81\%                  & 78\%                  & 109                   & 107                   & 106                   & 105                   \\ \bottomrule
\end{tabular}

\end{table}

\begin{table}

\caption{\label{tab:PB30}Percentage of patients on the correct treatment for them, and average total observed successes using CARA FLGI with $B=2$. True treatment difference of 30\%.}
\begin{tabular}{@{}lllllllll@{}}
\toprule
               & \multicolumn{4}{c}{\textbf{Patients on Correct Treatment}}                                    & \multicolumn{4}{c}{\textbf{Average Total Successes}}                                          \\ \midrule
               & \textbf{$n_z$=1} & \textbf{$n_z$=2} & \textbf{$n_z$=3} & \textbf{$n_z$=4} & \textbf{$n_z$=1} & \textbf{$n_z$=2} & \textbf{$n_z$=3} & \textbf{$n_z$=4} \\
\textbf{N=40}  & 86\%                  & 80\%                  & 76\%                  & 73\%                  
& 30                    & 30                    & 29                    & 29                    \\\textbf{N=80}  & 92\%                  & 87\%                  & 84\%                  & 81\%                  & 62                    & 61                    & 60                    & 59                    \\
\textbf{N=160} & 96\%                  & 93\%                  & 90\%                  & 88\%                   & 126                   & 124                   & 123                   & 122                   \\\bottomrule
\end{tabular}

\end{table}

These results are especially promising when considering the amount of patient benefit that the adaptive randomization offers. Tables \ref{tab:PB20} - \ref{tab:PB30} show the percentage of patients that were on the correct treatment for them across the simulations, for a true underlying treatment difference of 20\% and 30\%.  Compared to the 50\% which occurs when equal randomization is implemented, this is a stark improvement. The tables also show the average total number of successes per trial. Again, there are unsurprisingly far more successes observed for the adaptive design.  With comparable power to equal randomization for small block sizes in practice, a clear indication that the proposed procedure has the potential for success. 

\subsection{Illustrative Multi-Arm Example} \label{sec:mot_ex}

In order to demonstrate the use of FLGI allocation probabilities to test for superiority in a multi-arm setting, we use the following trial reported by \cite{Attarian2014} which looked at the a combination of baclofen, naltrexone and sorbitol (PXT3003) in patients with Charcot-Marie-Tooth disease type 1A as an illustrative example. A total of 80 patients were randomized to either the experimental treatment PXT3003 in 3 different doses, or a control group receiving a placebo. In this trial, equal randomization was used, with 19 patients randomized to the control group, and 21, 21 and 19 patents were allocated to the low, intermediate and high doses of PXT3003 respectively. The aim of the study was to assess both safety and tolerability as well as efficacy, with the measure of safety and tolerability the total number of adverse events. In the placebo group, a total of 9 out of 19 patients suffered adverse events, whereas in the active treatment groups this was 5, 7 and 6 in the low, intermediate and high dose groups respectively. 

We will use this example to simulate how this four-armed trial with only a single biomarker category would have looked using the FLGI procedure with a block size of 2. As before, we discard the first two allocation probabilities as burn in and test if the allocation probability exceeds 25\% across blocks 3 to 40. As there is only one category, we find the critical value, $c_q$ to be 30. In this example, we consider only pairwise comparisons between individual active treatments and control each at full level $\alpha$ for simplicity. Should overall control of the family-wise error rate be desired, standard approaches such as a Bonferroni correction or similar adjustment \citep{Simes1986} can be applied.

We will consider three scenarios of varying success rates across the four arms in this illustration. In the first scenario all treatments (including control) have the same success rate of 0.5, while the second scenario uses the estimated success rates from the study itself. The final situation considers a linear dose-response relationship from 0.53 to 0.77 (the lowest to highest observed success rates in the trial) across the four treatment arms.

In 10,000 replications of the trial under the null hypothesis, the type I error rate was well controlled at 5\% for the procedure using FLGI allocation probabilities, but was conservative for the Fisher exact test, for both allocation schemes. In scenario 2 which mimicked the results of the study (i.e. some difference between control and active, but hardly any difference between the different doses) on average 38 patients were allocated to the active treatment group with the best underlying success rate and the average total number of successes was 56 compared to the 53 in the original trial. In 34\% of the simulations the null hypothesis could be rejected using allocation probabilities compared to 32\% to using the original randomization scheme and a Fisher exact test.

In the final scenario of a linear dose-response a similar trend was observed. Using allocation probabilities to test for superiority led to an increase in power to 41\% from the 35\% of the Fisher exact test applied to the observed successes in the trial with equal randomization. 

\section{Discussion\label{sec:discussion}}
In this paper we have introduced an alternative inference approach to analyse a clinical trial that has been conducted using an FLGI design. This novel approach uses the allocation probabilities generated in the FLGI procedure as opposed to a test based on observed success/failure outcomes, and in doing so addresses the low power associated with unequal sample sizes inherent in response adaptive designs.

Although in this paper we have shown promising results for trials implementing the CARA FLGI rule for patient allocation, it is widely applicable in trials utilising any FLGI rule. Such is the focus here since it is shown to be near-optimal in terms of patient benefit \citep{Villar2015b}. In fact, such is the generality of this novel approach for inference that it can be applied to trials using other response adaptive randomized allocation rules that do not target a specific allocation, for example a ``play the winner'' style rule \citep[see][]{Wei1978}. We expect the approach to work best when the underlying response adaptive randomization deviates allocations significantly under the presence of a signal, like the FLGI rule does.

A standard approach in response-adaptive multi-arm trials to overcome the low power in such studies is to preserve the sample size of the control group \citep[e.g.][]{Trippa2012,Villar2015b}. This is either achieved by starting the trial with an initial period of equal allocation before applying response adaptive randomization or simply having a fixed allocation to the control arm throughout (although the latter is not applicable in the two arm case, which this novel testing approach is). Both of these do however reduce the patient benefit. One additional advantage of the novel testing approach that such arbitrary rules do not have is that testing on the basis of allocation probabilities yields good power without sacrificing on the patient benefit of using response adaptive randomization for the entire length of the study.

The only valid alternative approach to analyse clinical trials with response adaptive randomization is randomization based inference, which is known to be robust but reduce power compared to naive approaches. Our approach is the first alternative to be tailor made to these designs that increases power compared to such naive, and additionally not valid, analysis options.

In this work we have focused our explorations to pairwise testing of two treatment groups. However, response adaptive randomization is known to perform well for trials with multiple arms \citep{Wason2014}. While we illustrate how a pairwise testing strategy can be applied in this setting further extensions to global tests in multi-armed trials are of interest. 

Finally it is worth pointing out that although focus has been on showing superiority of a treatment in a pairwise test, the procedure can be adapted for two-sided tests by considering both tails of the null distribution as usual.

\section*{Acknowledgements}

 We are grateful for the discussions with Bie Verbist, An Vandebosch, Lixia Pei and Kevin Liu. The second author thanks the funding received from the National Institute for Health Research Cambridge Biomedical Research Centre at the Cambridge University Hospitals NHS Foundation Trust and the UK Medical Research Council (grant numbers: MC-UU-00002/3 and MC-UU-00002/14).  This report is independent research arising in part from Prof Jaki's Senior Research Fellowship (NIHR-SRF-2015-08-001) supported by the National Institute for Health and Social Care Research. The views expressed in this publication are those of the authors and not necessarily those of the NHS, the National Institute for Health Research or the Department of Health.\vspace*{-8pt}

 \bibliographystyle{biom} 
 \bibliography{library.bib}

\newpage

\section*{Supporting Information}
\section{WEB APPENDIX A: Additional Simulation Results}
\subsection{Web Figure 1}
\vspace{-20pt}
\begin{figure}[ht!]
\begin{center}
\includegraphics[width=0.7\textwidth]{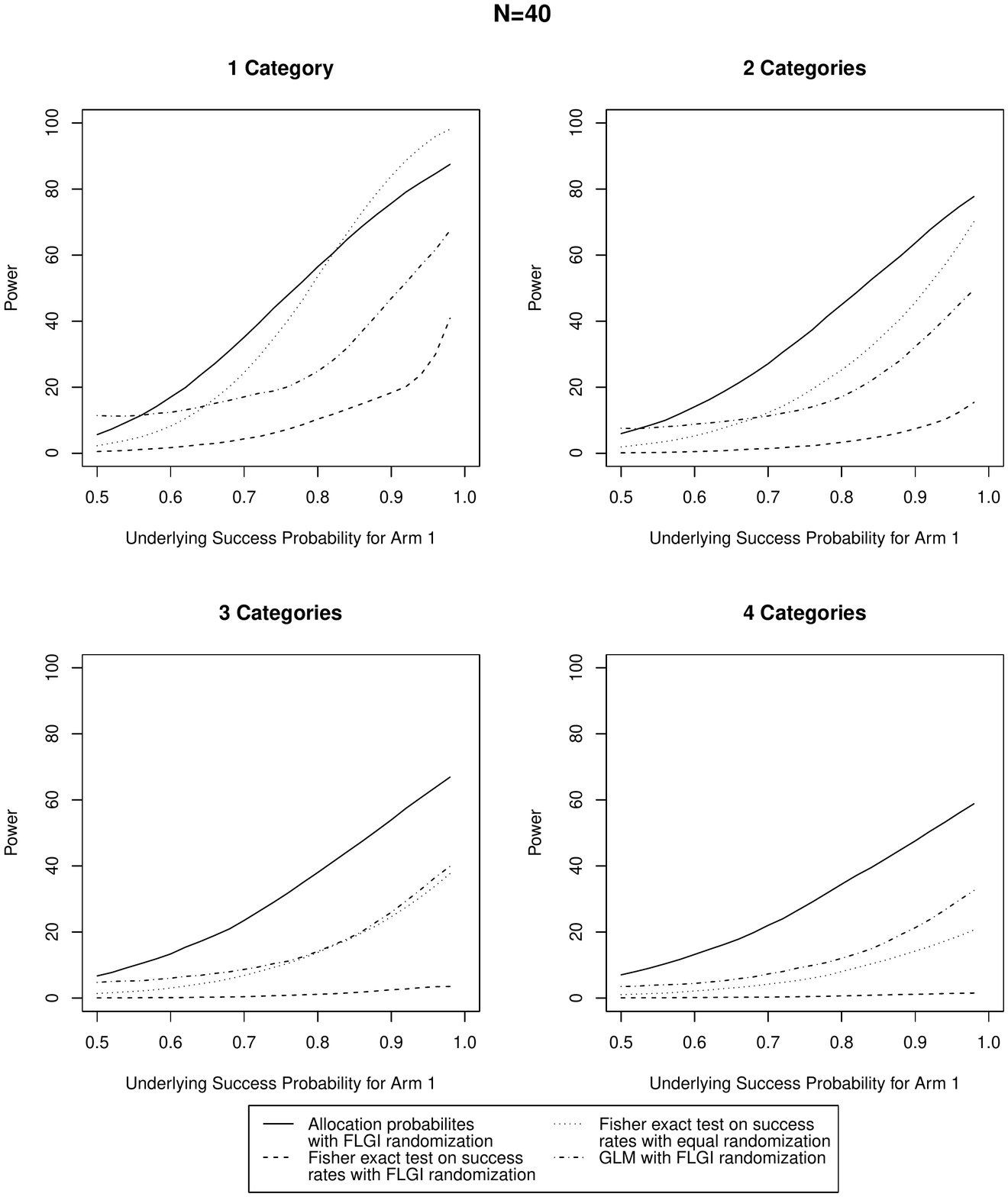}
\end{center}
\caption{Comparison of power for $N=40$ \& $B=2$.}
\label{fig:N40}
\end{figure}
\newpage
\subsection{Web Figure 2}
\begin{figure}[ht!]
\begin{center}
\includegraphics[width=0.7\textwidth]{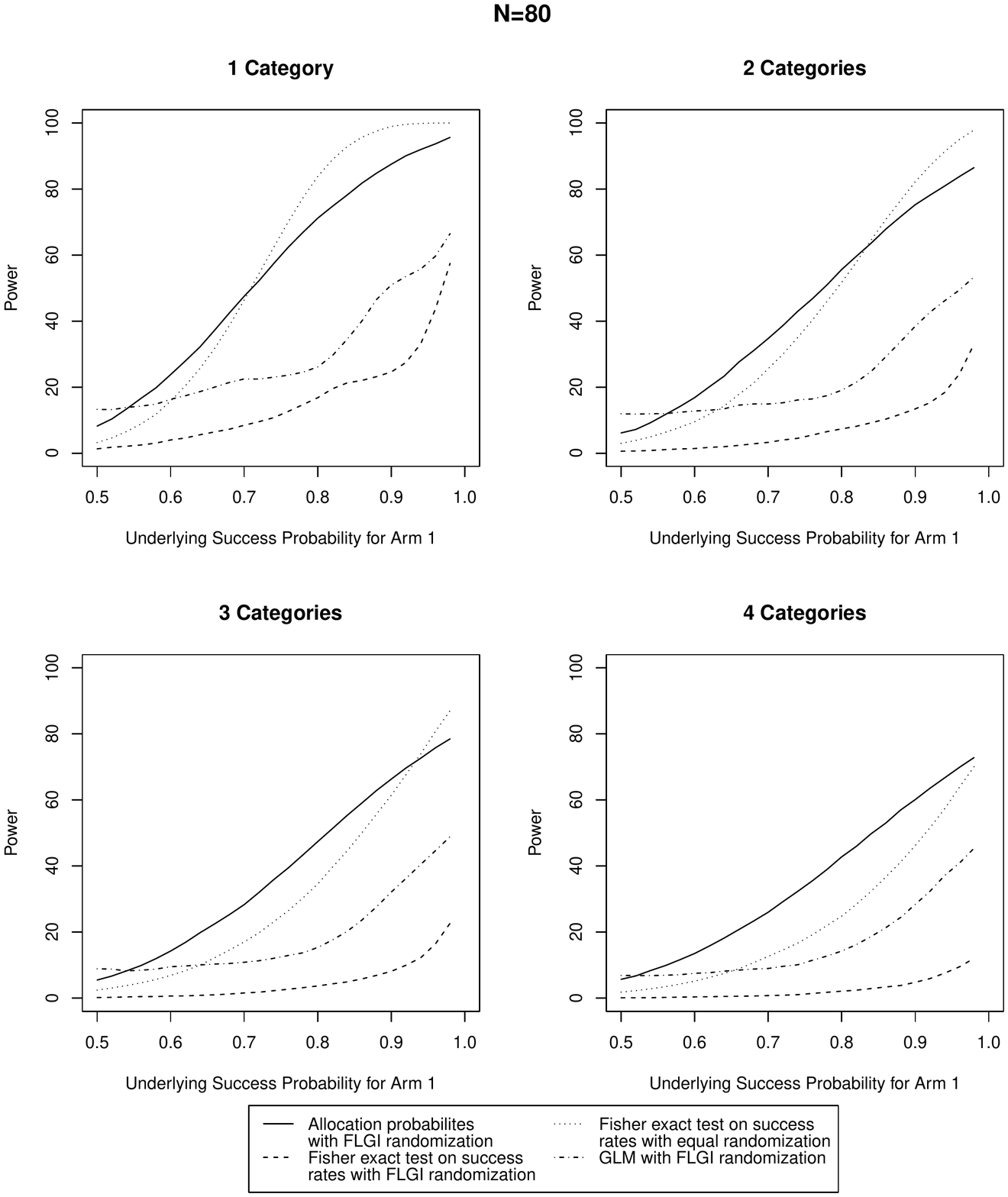}
\end{center}
\caption{Comparison of power for $N=80$ \& $B=2$.}
\label{fig:N80}
\end{figure}
\newpage

\subsection{Web Figure 3}
\begin{figure}[ht!]
\begin{center}
\includegraphics[width=0.7\textwidth]{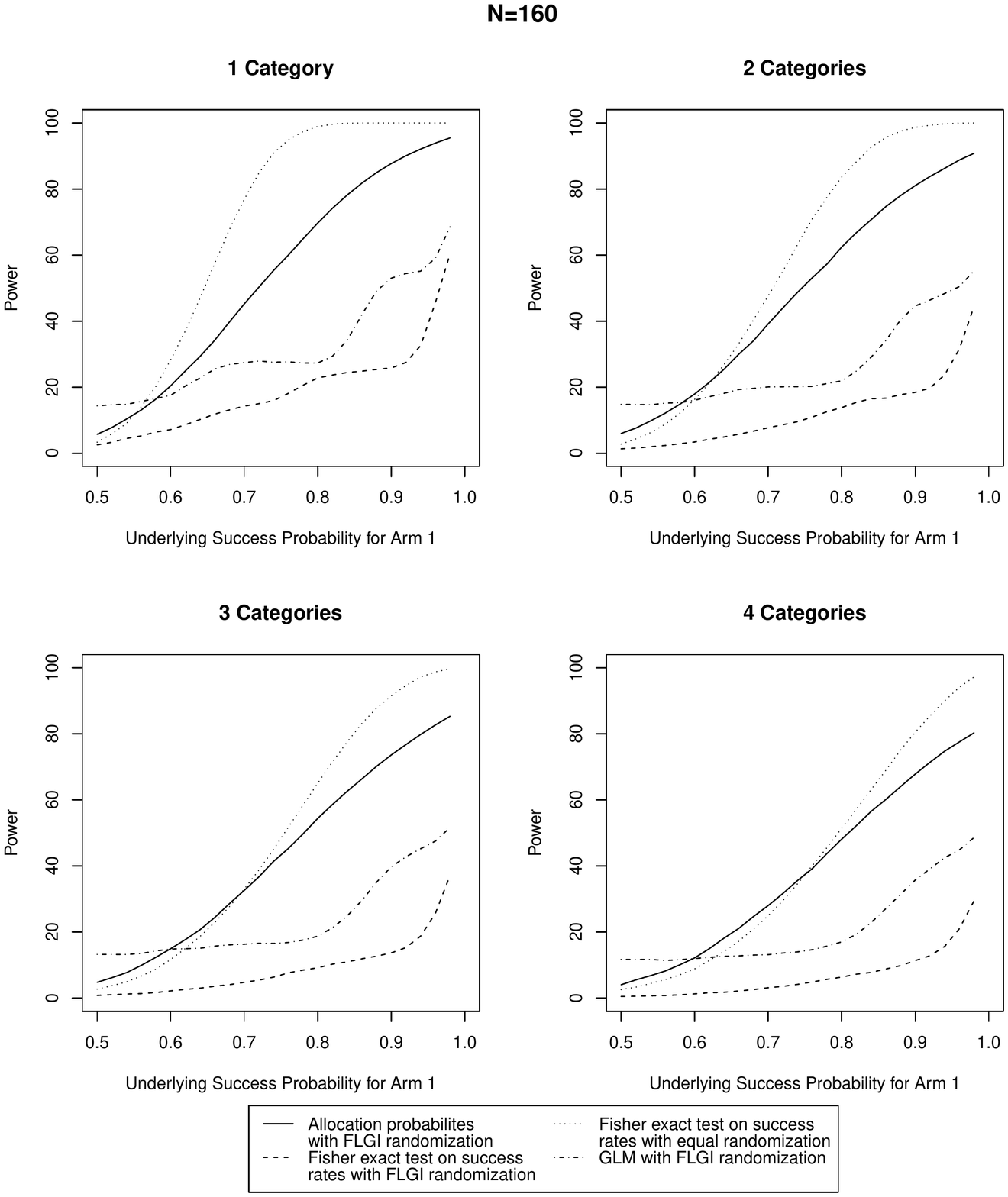}
\end{center}
\caption{Comparison of power for $N=160$ \& $B=2$.}
\label{fig:N160}
\end{figure}
\newpage
\subsection{Web Figure 4}
\begin{figure}[ht!]
\begin{center}
\includegraphics[width=0.7\textwidth]{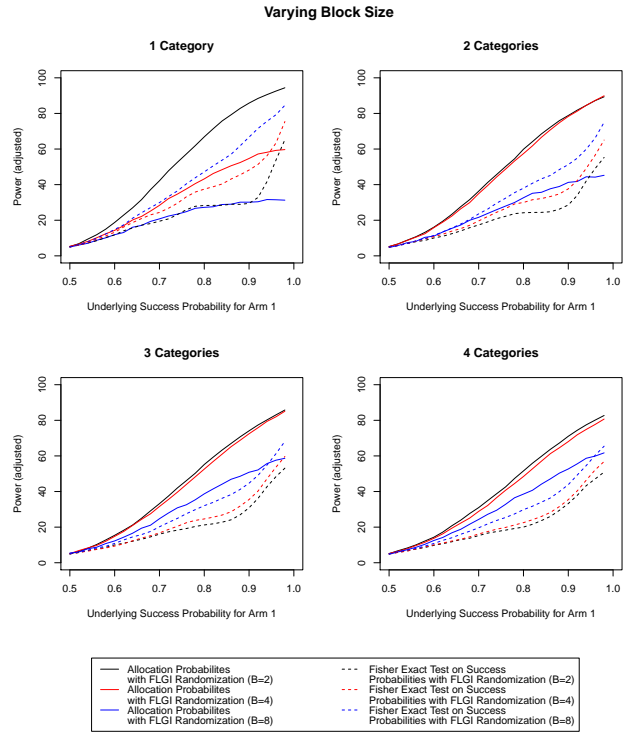}
\end{center}
\caption{Comparison of power of procedure using allocation probabilities vs Fisher exact test on success rates for $N=160$ with $B=2,4,8$; rejection criteria adjusted for type I error rate.}
\label{fig:block_size}
\end{figure}

\newpage
\section{WEB APPENDIX B: Derivation of Joint Distribution of $X_{j,z}$ \& $Y_{j,z}$} \label{app:XY}
This is a discrete distribution described by a $(B +1) \times (B +1)$ matrix of probabilities of observing each of the pairs of values of $X_{j,z}$ and $Y_{j,z}$ from 0 to $B$.

We first calculate the distribution of $Y_{j,z}$ conditional on $X_{j,z}$ by essentially following a tree diagram starting where the origin is the starting state. This makes decisions using the $GI$, choosing the arm with the highest $GI$ and the success probability on that arm is the posterior success probability so far on that arm. When the $GI$ is equal, the tree splits into two alternative routes with equal probability, as the algorithm would choose between the arms at random.

We take a ``snapshot'' after each value of $X_{j,z}$, telling us the probability of each outcome conditional on there being $X_{j,z}=x_{j,z}$ number of patients belonging to category $z$ within the block. To calculate the joint distribution, we multiply this conditional probability by the probability of there being $x_{j,z}$ number of patients in category $z$, following the Binomial distribution previously introduced.

\begin{figure}[ht!]
\begin{center}
\includegraphics[width=0.8\textwidth]{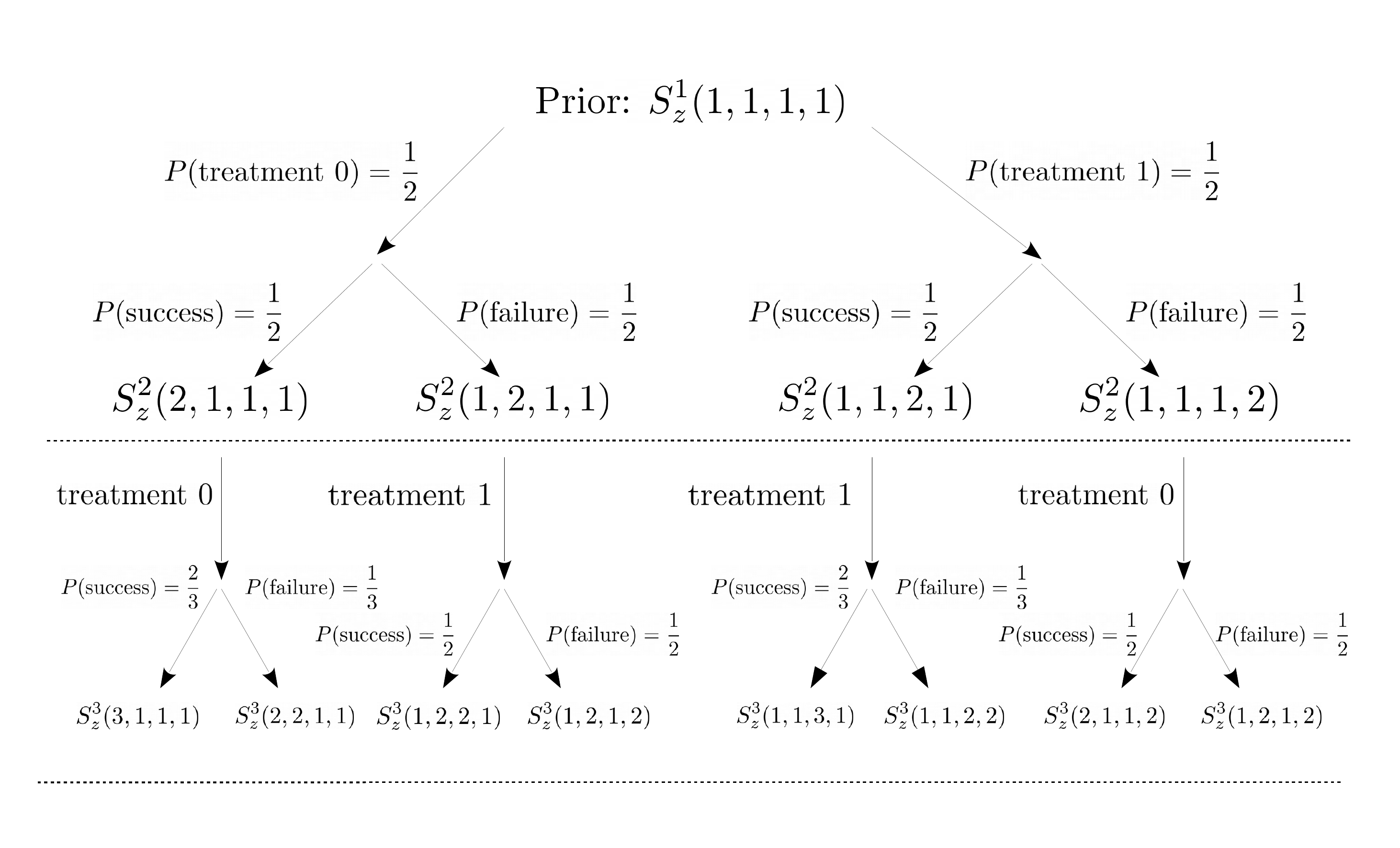}
\end{center}
\caption{An illustration of tree diagram used in calculation of distribution of $p_{\mbox{{\small alpro, z}}}$, with $B=2$.}
		\label{fig:allo_tree}
\end{figure}


As an illustration, Figure \ref{fig:allo_tree} shows how this is calculated with a block size of 2. The probabilities on the tree are all conditional on $X_{j,z}=2$, implying that of the block of two patients, both are in the given biomarker category $z$. Success and failure probabilities are the posterior probabilities for a given treatment . The dashed line indicates the level at which to take each ``snapshot''. For example the probability of having two patients in category $z$ allocated to treatment 1 conditional on two patients being in category $z$ is the sum of following the two branches that result in state $S^3_z(1,1,3,1)$ and $S^3_z(1,1,2,2)$. These branches have probability $\frac{1}{2}\times\frac{1}{2}\times\frac{2}{3}=\frac{1}{6}$ and $\frac{1}{2}\times\frac{1}{2}\times\frac{1}{3}=\frac{1}{12}$ respectively, and hence the $\mathds{P}(Y_{j,z}=2|X_{j,z}=2)=\frac{1}{4}$ and therefore $\mathds{P}(Y_{j,z}=2 \hspace{5pt} \& \hspace{5pt} X_{j,z}=2)=\mathds{P}(Y_{j,z}=2|X_{j,z}=2)\mathds{P}(X_{j,z}=2)=\frac{1}{4n^2_{z}}$.\\

In the general case, the joint distribution of $Y_{j,z}$ and $X_{j,z}$ is given as:
\begin{align*}
\mathds{P} (Y_{j,z} = y_{j,z}\ \&\ X_{j,z}=x_{j,z}) = \binom{B}{x_{j,z}}& \left( \frac{1}{n_{z}} \right)^{x_{j,z}} \left( \frac{1-n_{z}}{n_{z}} \right)^{B-x_{j,z}} \times \\ &\mathds{P} (Y_{j,z} = y_{j,z} | X_{j,z}=x_{j,z}) \\
=  \binom{B}{x_{j,z}} &\left( \frac{1}{n_{z}} \right)^{x_{j,z}} \left( \frac{1-n_{z}}{n_{z}} \right)^{B-x_{j,z}} \times \\ & \sum_{\mathcal{S}^i_{z,x_{j,z},y_{j,z}}} \mathds{P} (S^i_{z}(s^i_{0,z},f^i_{0,z},s^i_{1,z},f^i_{1,z})),
\end{align*}\normalsize
where $\mathcal{S}^i_{z,x_{j,z},y_{j,z}}$ is the set of all states $S^i_{z} = (s^i_{0,z},f^i_{0,z},s^i_{1,z},f^i_{1,z})$ such that $s^i_{0,z} + f^i_{0,z} + s^i_{1,z} + f^i_{1,z} - (s^1_{0,z}+f^1_{0,z}+s^1_{1,z}+f^1_{1,z}) = x_{j,z}$ and $s^i_{1,z}+f^i_{1,z} - (s^1_{1,z}+f^1_{1,z}) = y_{j,z}$ . 

For the generic state $S^i_z(s^i_{0,z},f^i_{0,z},s^i_{1,z},f^i_{1,z})$, let $\mathfrak{S}^{i-1}_z$ be the set of four possible different states that may precede this state. Then 
\[
\mathds{P} (S^i_z(s^i_{0,z},f^i_{0,z},s^i_{1,z},f^i_{1,z}))= \sum_{S^{i-1}_z \in \mathfrak{S}^{i-1}_z} \mathds{P} (S^i_z(s^i_{0,z},f^i_{0,z},s^i_{1,z},f^i_{1,z})|S^{i-1}_z)\mathds{P}(S^{i-1}_z)
\]

Let $\kappa^{i-1}_z$ equal 1 when the previous patient in category $z$ was allocated to the experimental treatment and 0 if allocated to the control. Then let $T^{i-1}_{\kappa^{i-1}_z,z}$ equal 1 if the outcome on treatment $\kappa^{i-1}_z$ was a success and 0 if it was a failure, so that for example
\[
\mathds{P} (\kappa^{i-1}_z=0, T^{i-1}_{0,z}=1) = \mathds{P} (S^i_z(s^i_{0,z},f^i_{0,z},s^i_{1,z},f^i_{1,z})|S^{i-1}_z(s^i_{0,z}-1,f^i_{0,z},s^i_{1,z},f^i_{1,z})).
\]

This is calculated for all four states by
\begin{equation*}
  \mathds{P} ((\kappa^{i-1}_z=\kappa, T^{i-1}_{\kappa,z}=t)) =
    \begin{cases}
      0 & \text{if ${\scriptstyle GI(s^i_{\kappa,z}-t,f^i_{\kappa,z}-(1-t)) < GI(s^i_{1-\kappa,z},f^i_{1-\kappa,z})}$}\\
      \frac{ts^i_{\kappa,z} +(1-t)f^i_{\kappa^{i-1},z}-1}{2(s^i_{\kappa,z}+fi_{\kappa,z}-1)} & \text{if $ {\scriptstyle GI(s^i_{\kappa,z}-t,f^i_{\kappa,z}-(1-t)) = GI(s^i_{1-\kappa,z},f^i_{1-\kappa,z})}$}\\
      \frac{ts^i_{\kappa,z} +(1-t)f^i_{\kappa,z}-1}{(s^i_{\kappa,z}+fi_{\kappa,z}-1)} & \text{if $ {\scriptstyle GI(s^i_{\kappa,z}-t,f^i_{\kappa,z}-(1-t)) > GI(s^i_{1-\kappa,z},f^i_{1-\kappa,z})}$},
    \end{cases}       
\end{equation*}

Each of the probabilities of the states $S_{z}^{i-1}$ can then be recursively calculated in the same way.
\newpage
\section{WEB APPENDIX C: Proofs} \label{app:proofs}
\subsection{Proof of Lemma 3.1}
\begin{Proof}
By symmetry, the distribution of allocation probability to the experimental treatment ($p_{\mbox{{\small alpro, z}}}^{\mbox{{\small $t_1$}}}$) at $S^i_z(s^i_{0,z}, f^i_{0,z}, s^i_{1,z}, f^i_{1,z})$ is the same as the distribution of the allocation probability to the standard treatment ($p_{\mbox{{\small alpro, z}}}^{\mbox{{\small $t_0$}}}$) at $S^i_z(s^i_{1,z}, f^i_{1,z},s^i_{0,z}, f^i_{0,z})$, regardless of assumptions on treatment difference. Since from any state $p_{\mbox{{\small alpro, z}}}^{\mbox{{\small $t_1$}}}=1-p_{\mbox{{\small alpro, z}}}^{\mbox{{\small $t_0$}}}$, we may take, for $0 \leq c \leq 1$
\begin{align*}
F_{s^i_{0,z} f^i_{0,z} s^i_{1,z} f^i_{1,z}}^{t_1}(c) &= F_{s^i_{1,z} f^i_{1,z} s^i_{0,z} f^i_{0,z}}^{t_0}(c) \\
&= \mathds{P}(p_{\mbox{{\small alpro}}, s^i_{1,z} f^i_{1,z}s^i_{0,z} f^i_{0,z}}^{\mbox{{\small $t_0$}}}<c) \\
&= \mathds{P}(1-p_{\mbox{{\small alpro}}, s^i_{1,z} f^i_{1,z}s^i_{0,z} f^i_{0,z}}^{\mbox{{\small $t_1$}}}<c) \\
&= \mathds{P}(1-c<p_{\mbox{{\small alpro}}, s^i_{1,z} f^i_{1,z}s^i_{0,z} f^i_{0,z}}^{\mbox{{\small $t_1$}}}) \\
&= 1- \mathds{P}(p_{\mbox{{\small alpro}}, s^i_{1,z} f^i_{1,z}s^i_{0,z} f^i_{0,z}}^{\mbox{{\small $t_1$}}}<1-c) \\
&= 1- F_{s^i_{1,z} f^i_{1,z}s^i_{0,z} f^i_{0,z}}^{t_1}(1-c).
\end{align*}

Therefore $F_{s^i_{0,z} f^i_{0,z}s^i_{1,z} f^i_{1,z}}^{t_1}(0.5)=1- F_{s^i_{1,z} f^i_{1,z}s^i_{0,z} f^i_{0,z}}^{t_1}(0.5)$.
\end{Proof}

\subsection{Proof of Lemma 3.2}
\begin{Proof}
We intend to prove this in two steps. First, we show that consistently higher allocation probabilities for one arm indicates a true treatment difference. We prove this by showing the contrapositive; that no true treatment difference implies the allocation probability is equally likely to be above or below 0.5 within a trial of size $N$. 

The distribution of $p_{\mbox{{\small alpro, z}}}^{\mbox{{\small $t_1$}}}$ at the beginning of block $k$ is given by the mixture distribution in equation (4) in the main paper. Consider the set of all states $Z$ as pairs of ``mirror'' states and individual ``symmetrical'' states (where $s^i_{0,z}=s^i_{1,z}$ \& $f^i_{0,z}=f^i_{1,z}$). Take first the ``symmetrical'' states, noting that $GI(s^i_{0,z},f^i_{0,z})=GI(s^i_{1,z},f^i_{1,z})$. We therefore have $\mu_{y,z}=\frac{\mu_{x,z}}{2}$, since within each block of the Monte-Carlo runs, we expect to see equal allocations between treatments. Substituting this into $F$, we see that $F(0.5)=0.5$, indicating that the $p_{\mbox{{\small alpro, z}}}$ for both treatments is equally likely to be above and below 0.5. Likewise this is true for the exact calculations by \citet{Villar2015b}.

Now consider the pairs of ``mirror'' states. As a consequence of lemma 3.1,  the distribution of the allocation probabilities to the experimental treatment from these pairs of states are reflections about $p_{\mbox{{\small alpro, z}}}=0.5$. Under the assumption that $p_1-p_0=0$, these states are also equally probable. Hence, when the weighted sum of distributions for both ``mirror'' and ``symmetrical'' states is calculated in the following Section (equation (4) in the main paper), the overall effect is that allocation probabilities above and below 0.5 are equally likely.

Now that we have established that consistently higher allocation probabilities for one arm imply $p_1 \neq p_0$, we consider the inequality relating $p_1$ and $p_0$ when a given arm has consistently higher allocation probabilities. This is a strict inequality, either $p_1>p_0$ or $p_1<p_0$ since in the first part of this proof we showed $p_1 \neq p_0$.

For the second part of this proof, we show that consistently higher allocation probabilities for arm 1 imply $p_1>p_0$. We assume for contradiction the negation of this statement, that we have consistently higher allocation probabilities for arm 1 and $p_1<p_0$. Since the allocation probabilities for arm 1 are consistently higher, more patients in the trial are allocated to treatment 1 than treatment 0. However, because the true success probability of treatment 1 is lower than treatment 0, this trial design has fewer expected success than a trial design where arm 0 has consistently higher allocation probabilities and hence more patients are allocated to treatment 0 than treatment 1. Since the FLGI procedure is near optimal in maximising expected successes, we have a contradiction in that there clearly exists another design with a much higher level of expected successes. Hence consistently higher allocation probabilities for arm 1 do indeed imply $p_1>p_0$.

\end{Proof}

\subsection{Proof of Lemma 3.3}
\begin{Proof}
As the number of patients in the trial, $N$, gets very large, the posterior success probabilities for each of the two treatments will tend to the true success probabilities for those treatments. Therefore, for any finite block size $B$ and true difference in success probabilities $p_1-p_0>0$, there exists $\epsilon >0$ such that

\[
\mathds{P} \left( \left| \frac{s^N_{1,z}}{s^N_{1,z} + f^N_{1,z}} - \frac{s^N_{0,z}}{s^N_{0,z} + f^N_{0,z}} - (p_1-p_0) \right| < \epsilon \right) \rightarrow 1 \hspace{10pt} \mbox{as} \hspace{10pt} N \rightarrow \infty.
\]

Hence for all $0 < m \leq 1$, when $p_1 -p_0 \geq m$ there exists $0<\epsilon < m$ such that
\[
\mathds{P} \left( \frac{s^N_{1,z}}{s^N_{1,z} + f^N_{1,z}} - \frac{s^N_{0,z}}{s^N_{0,z} + f^N_{0,z}} \geq m - \epsilon \right) \rightarrow 1 \hspace{10pt} \mbox{as} \hspace{10pt} N \rightarrow \infty.
\]

Therefore for all $b=0,\ldots,B$ 
\[
\mathds{P} \left( GI(s^N_{1,z}, f^N_{1,z} + b ) > GI(s^N_{0,z},f^N_{0,z}) \right) \rightarrow 1 \hspace{10pt} N \rightarrow \infty.
\]

In practice this means that in the state $S^N_z(s^N_{0,z}, f^N_{0,z}, s^N_{1,z}, f^N_{1,z})$, no matter how unlikely the outcome (success/failure) and category classification (all/no patients in given category), every patient in the following block in the Monte-Carlo simulations used to calculate the allocation probability will be allocated to the experimental treatment. Consequently, in equation (1) in the main paper, we have $X_{j,z}=Y_{j,z}$ for all $j=1, \ldots, n$ and hence $p^{t_1}_{\mbox{{\small alpro, z}}}=1$.

In fact, as long as $N>>B$, the state $S^N_z(s^N_{0,z}, f^N_{0,z}, s^N_{1,z}, f^N_{1,z})$ with $GI(s^i_{0,z},f^i_{0,z})<GI(s^i_{1,z},f^i_{1,z})$ is more likely than its ``mirror'' state $S^N_z( s^N_{1,z}, f^N_{1,z}, s^N_{0,z}, f^N_{0,z})$ when $p_1-p_0>m$. Further, as a consequence of lemma 3.1, the distribution of the allocation probabilities to the experimental treatment from these states are reflections about $p_{\mbox{{\small alpro, z}}}=0.5.$ Therefore when the weighted sum of all potential distributions is calculated in the following Section (equation (4) in the main paper), the overall effect is that $\mathds{P}(p_{\mbox{{\small alpro, z}}}>0.5)>0.5$.

\end{Proof}

\newpage
\section{WEB APPENDIX D: Specification of null distribution} \label{sec:app_null}

For a block of size $B$, there are $\eta= 0,1, \ldots B$ potential patients in the block in a given biomarker category $z$. Therefore the number of potential states at the end of block 1 is:

\[
M_1=\sum_{{\eta}=0}^{B} \binom{{\eta} + 3 }{3}.
\]
Since in general, the number of ways to put $r$ objects in $R$ boxes is $\binom{r+R-1}{R-1}$. For a given ${\eta}$, we have ${\eta}$ objects (patients), to put in 4 boxes ($s_{0,z}$, $f_{0,z}$, $s_{1,z}$, $f_{1,z}$). $M_1$ is therefore the sum over the number of potential patients within the category and block, of the number of potential states at the end of block 1. 

Thus for the next blocks, the number of potential states at the end of block $k$ is
\[
M_{k}=\sum_{{\eta}=0}^{kB}\binom{{\eta} + 3}{3}.
\]

Define $f_{\zeta}$ as the distribution of allocation probabilities from a given state ${\zeta}$, as described in equation (3) in the main paper. Let $c_p$ be the previous allocation probability, so in the following when integrating over [0.5,1], it indicates we are conditioning on a previous $\alpha_k=1$ whereas integrating over [0,0.5] indicates we are conditioning on a previous $\alpha_k=0$.

Let $\mathds{P} ({\zeta} |c_p) $ be the probability of being in state ${\zeta}$ conditional on the previous allocation probability. That is, the allocation probability that was calculated at the beginning of the previous block. This can be calculated in a similar manner as shown in Figure \ref{fig:allo_tree}. A tree starting at each potential state the previous block could have ended in is needed, following the possible allocations and success/failure along the branches to find the conditional probability. Again including a multiplicative factor of the relevant binomial probability at each dotted line to calculate the joint probability. Two major differences however, are that patient allocation is random with probability determined by the previous calculated allocation probability $c_p$ as opposed to the deterministic $GI$ rule, and the success is determined by a pre-specified equal success probability for both treatments as opposed to posterior probability.

With an uninformative prior, for the first block we simply have the case that $\mathds{P}(\alpha_1=0)=\mathds{P}(\alpha_1=1)=0.5$. Then for each potential state ${\zeta}$ that the first block could end in, the joint probability of the first block ending in that state ${\zeta}$ and the first allocation probability being less than 0.5 is

\begin{equation}
\mathds{P} ({\zeta}, \alpha_1=0)= \int_{c_p=0}^{c_p=0.5} \mathds{P}({\zeta}|prior,c_p) f_{prior}(c_p).
\end{equation}

We define the set of all states ${\zeta}$ such that $\mathds{P} ({\zeta},\alpha_1=0) \neq 0$ as $Z_0$, and the set of all states ${\zeta}$ such that $\mathds{P} ({\zeta},\alpha_1=1) \neq 0$ as $Z_1$

These satisfy the following
\begin{equation}
\mathds{P} (\alpha_1=0)= \sum_{{\zeta} \in Z_0} \int_{c_p=0}^{c_p=0.5} \mathds{P}({\zeta}|prior,c_p) f_{prior}(c_p)dc_p=0.5,
\end{equation}
and
\begin{equation}
\mathds{P} (\alpha_1=1)= \sum_{{\zeta} \in Z_1} \int_{c_p=0.5}^{c_p=1} \mathds{P}({\zeta}|prior,c_p) f_{prior}(c_p)dc_p=0.5.
\end{equation}

Then consider the second block. Again, the joint probability of the second block ending in state ${\zeta}$ and the first two allocation probabilities being less than 0.5 is
\begin{equation}
\mathds{P} ({\zeta},\alpha_1=0,\alpha_2=0)= \sum_{{\zeta}_p \in Z_0} \mathds{P}({\zeta}_p,\alpha_1=0) \int_{c_p=0}^{c_p=0.5} \mathds{P}({\zeta}|{\zeta}_p,c_p) f_{{\zeta}_p}(c_p)dc_p.
\end{equation}

We define the set of all states ${\zeta}$ such that $\mathds{P} ({\zeta},\alpha_1=0,\alpha_2=0) \neq 0$ as $Z_{00}$, and in general $Z_{\alpha_1 \alpha_2}$ is the set of all states ${\zeta}$ that have a non-zero probability of occurrence at the end of block 2 when the first two allocation probabilities are less than 0.5.

So that we can for example calculate the probability of the first two allocation probabilities being less than 0.5 by summing over the set $Z_{00}$
\begin{equation}
\mathds{P} (\alpha_1=0,\alpha_2=0)=\sum_{{\zeta} \in Z_{00}} \mathds{P} ({\zeta},\alpha_1=0,\alpha_2=0)
\end{equation}

The calculations for subsequent blocks follow on, for example if we wish to calculate the probability of the sequence $\alpha_1=0,\alpha_2=0,\alpha_3=1$, we first calculate
\begin{equation}
\mathds{P} ({\zeta},\alpha_1=0,\alpha_2=0,\alpha_3=1)= \sum_{{\zeta}_p \in Z_00} \mathds{P}({\zeta}_p,\alpha_1=0,\alpha_2=0) \int_{c_p=0.5}^{c_p=1} \mathds{P}({\zeta}|{\zeta}_p,c_p) f_{{\zeta}_p}(c_p)dc_p.
\end{equation}
Then sum over the set $Z_{001}$
\begin{equation}
\mathds{P} (\alpha_1=0,\alpha_2=0,\alpha_3=1)=\sum_{{\zeta} \in Z_{001}} \mathds{P} ({\zeta},\alpha_1=0,\alpha_2=0,\alpha_3=1).
\end{equation}

However, these calculations are extremely computationally intensive once larger numbers of blocks are considered, and hence a very close approximation may be used. This approximation no longer integrates over $c_p$ but instead finds the expectation of $c_p$ with respect to the distribution $f_{{\zeta}_p}$, $\mathds{E}_{f_{{\zeta}_p}} [c_p]$ and approximates in the following way:

\begin{equation}
\mathds{P} ({\zeta},\alpha_1=0,\alpha_2=0)= \sum_{{\zeta}_p \in Z_0} \mathds{P}({\zeta}_p,\alpha_1=0) \mathds{P}({\zeta}|{\zeta}_p,c_p=\mathds{E}_{f_{{\zeta}_p}} [c_p]) .
\end{equation}

Due to the nature of the procedure, the ordering of the $\alpha_k$ is not exchangeable. For example $\mathds{P}(\alpha_1=1, \alpha_2=1, \alpha_3=0)$ is not necessarily the same as $\mathds{P}(\alpha_1=1, \alpha_2=0, \alpha_3=1)$, even though for both sequences $\sum_{k=1}^{3}\alpha_k=2$. To calculate the probability of observing a given value for $\sum_{k=1}^{K}\alpha_k$, the probability of each sequence with this corresponding value of $\sum_{k=1}^{K}\alpha_k$ is summed over, since they are mutually exclusive.

From this discrete null distribution, we can find a critical value for 5\% type I error in the test.




\label{lastpage}

\end{document}